 \documentclass[journal,article,submit,moreauthors,dvipdfm,12pt,a4paper]{mdpi} 
 \usepackage{graphicx}
 \usepackage{epsfig}
\lastpage{x}
\doinum{10.3390/------}
\pubvolume{xx}
\pubyear{2013}
\history{Received: xx / Accepted: xx / Published: xx}
\newcommand{\arcsec}{\hbox{$^{\prime\prime}$}}

\def\gapprox{\lower.4ex\hbox{$\;\buildrel >\over{\scriptstyle\sim}\;$}}
\def\lapprox{\lower.4ex\hbox{$\;\buildrel <\over{\scriptstyle\sim}\;$}}
\def\ang{\AA}
\Title{Optimization of Curvi-Linear Tracing Applied to Solar Physics 
and Biophysics}

\Author{Markus J. Aschwanden $^{1*}$, Bart De Pontieu $^{1}$, and 
	Eugene A. Katrukha $^{2}$ }

\address{%
$^{1}$ Lockheed Martin Solar and Astrophysics Laboratory, Bldg. 252, Org. A021S,
3251 Hanover St., Palo Alto, CA 94304, USA \\
$^{2}$ Cell Biology, Faculty of Science, Utrecht University, Padualaan 8, 
3584CH, The Netherlands }

\corres{aschwanden@lmsal.com, Phone: 650-424-4001, Fax: 650-424-3994.}

\abstract{We developed an automated pattern recognition code that is
particularly well suited to extract one-dimensional curvi-linear features
from two-dimensional digital images. A former version of this 
{\sl Oriented Coronal CUrved Loop Tracing (OCCULT)} code was applied
to spacecraft images of magnetic loops in the solar corona, recorded 
with the NASA spacecraft {\sl Transition Region And Coronal Explorer (TRACE)} 
in extreme ultra-violet wavelengths. Here we apply an advanced version of this
code ({\sl OCCULT-2}) also to similar images from the {\sl Solar Dynamics
Observatory (SDO)}, to chromospheric H-$\alpha$ images obtained with the
{\sl Swedish Solar Telescope (SST)}, and to microscopy images of 
microtubule filaments in live cells in biophysics. We provide a full 
analytical description of the code, optimize the control parameters, 
and compare the automated tracing with visual/manual methods.
The traced structures differ by up to 16 orders of magnitude in size, 
which demonstrates the universality of the tracing algorithm.} 


\begin{document}

\section{Introduction}

Image segmentation is an image processing method that subdivides
an image into its constituent regions or objects, which can have 
the one-dimensional geometry of curvi-linear (1D) segments, or the
two-dimensional (2D) geometry of (fractal) areas. 
Common techniques include point, 
line, and edge detection, edge linking and boundary detection, Hough
transform, thresholding, region-based segmentation, morphological
watersheds, etc. (e.g., \cite{gonzales2008}). Since there exists no 
omni-potent automated pattern recognition code that works for all
types of images equally well, we have to customize suitable 
algorithms for each data type individually by taking advantage
of the particular geometry of the features of interest, using
{\sl a priori} information from the data. In this study we optimize
an automated pattern recognition code to extract magnetized
loops from images of the solar corona with the aim of optimum 
completeness and fidelity. We will demonstrate that the same code 
works also equally well for microscopic images in biophysics. 
The particular geometric property of
the extracted features is the relatively large curvature radius
of coronal magnetic field lines, which generally do not have sharp 
kinks and corners, but exhibit continuity in the variation of
the local curvature radius along their length. Using a related 
strategy of curvature constraints, coronal loops were extracted also 
with the directional 2D Morlet wavelet transform (\cite{biskri2010}).
In addition, solar coronal loops, as well as biological microtubule 
filaments, have a relatively small cross-section compared with their 
length, so that they can be treated as curvi-linear 1D objects in a
tracing method. Tracing of 1D structures with large curvature
radii simplifies an automated algorithm enormously, compared with
segmentation of 2D regions with arbitrary geometry and possibly
fractal fine structure \cite{mcateer2005}.

The content of this paper includes a brief description of the
automated tracing code (Section 2), applications to images in
solar physics (Section 3), to images in biophysics (Section 4),
discussion and conclusions (Section 5), and a full analytical
description of the code in Appendix A. 

\section{Description of the Automated Tracing Code}

Early versions of {\sl Oriented-Connectivity Methods (OCM)} applied
to solar images were pioneered by Lee, Newman, and Gary \cite{lee2004,lee2006}. 
This code was applied to a {\sl Transition Region And Coronal Explorer
(TRACE)} image and a total of 57 coronal loops were detected in 
a solar active region, which supposedly outline the dipolar magnetic field.
The results of this code were compared among five 
numerical codes based on similar curvi-linear loop-segmentation methods
in a study of Aschwanden \cite{aschwanden2008}, in terms of the
cumulative size distribution of loop lengths, the median and maximum
detected loop lengths, the completeness and detection efficiency, 
accuracy, and flux sensitivity. One of these codes was developed
further, which we call {\sl Oriented Coronal CUrved Loop Tracing 
(OCCULT)} code, and was found to approach the quality of visually 
and manually traced loops, detecting a total of 272 loop structures 
in the same TRACE image \cite{aschwanden2010}.
In the study here we developed this code further and tested it in a larger 
parameter space and with different types of images. Technical 
details of the original OCCULT code are given in \cite{aschwanden2010},
a concise description of the advanced OCCULT-2 code is given in the
following, while a comprehensive analytical description of OCCULT-2
is provided in Appendix A.

\begin{enumerate}

\item{\underbar{Background supression:} The median $z_{med}$
of an intensity image $z_{ij}=I_0(x_i,y_i)$ is computed, and the low 
intensity values with $z_{ij} < z_{min}$ are set to the base value $z_{min} =
z_{med} \times q_{med}$, with $q_{med}$ being a selectable control parameter,
with a default value $q_{med}=1.0$ if applied, and $q_{med}=0$ if ignored,
while a range of $1 \lapprox q_{med} \lapprox 2.5$ was found to be useful
for noisy data. The median value $z_{med}$ is 
a good estimate of the background (if the features of interest cover
less than 50\% of the image area) and can manually be adjusted with the
multiplier $q_{med}$ otherwise. The parts of the original image that have
intensities below this base level, are then rendered with a constant value, 
are noise-free, and will automatically suppress any structure detection 
in the background below this base level. This new method is more flexible
and efficient in suppressing faint background structures.}

\item{\underbar{Highpass and lowpass filtering:} A lowpass
filter with a boxcar smoothing constant $n_{sm1}$ smoothes out the
data noise (e.g., photon noise with Poisson statistics in astrophysical
and microscopy images, e.g., see \cite{pawley2008}),
while a highpass filter with a boxcar smoothing constant 
$n_{sm2}$ enhances the fine structure. The two combined filters 
represent a bandpass filter (with $n_{sm1} < n_{sm2}$), defined by 
\begin{equation}
	I^{filter}(x_i,y_j)= smooth[I_0(x_i,y_j),n_{sm1}]
	            - smooth[I_0(x_i,y_j),n_{sm2}] \ .
\end{equation}
For theoretical and experimental reasons, a filter combination
of $n_{sm2} = n_{sm1}+2$ yields the sharpest enhancement of the
intermediate spatial scale, and thus warrants optimum 
performance in tracing of curvi-linear structures.}

\item{\underbar{Initialization of loop structures:} The code initializes
the first structure to be traced from the position $(x_0^a, y_0^a)$ with 
the maximum brightness or flux intensity $f_0^a=I_0(x_0^a, y_0^a)$ in the 
original image $I_0(x,y)$. Once the full loop has been traced, the area of 
the detected loop is erased to zero, and the next loop structure is initialized 
at the position $(x_0^b, y_0^b)$ with the next flux maximum $f_0^b$ in the 
residual image. The initialization of subsequent loops $(f_0^c, f_0^d, ...)$
is continued iteratively until the residual image becomes entirely zeroed out,
the increase of detected structures stagnates, or a maximum loop number 
$N_{max}$ is reached.}
 
\item{\underbar{Loop structure tracing:} An initialized structure
starting at its flux maximum position $(x_0, y_0)$ is then traced 
in forward direction to the first end point of the loop, and 
then in opposite direction from the original starting point to the 
second endpoint. The two bi-directional segments are then combined 
into a single uni-directional 1D path $s_i=s(x_i, y_i)$, $i=1,...,n_s$. 
The step-wise tracing along a loop structure position $s_i$ is carried 
out by determining first the direction of the local ridge (defined by the 
azimuthal angle $\alpha_l$ with respect to the $x$-axis), and secondly
by determining the local curvature radius $r_m$. The curved segment that
follows a local ridge closest, is used as a second-order polynomial
to extrapolate the traced loop segment by one incremental step (of
$\Delta s=1$ pixel). This second-order guiding criterion represents
an improvement over the first-order guiding criterion used in the previous
OCCULT code \cite{aschwanden2010}. The second-order guiding criterion 
is defined by the brightness distribution $f(s)=f[x_{seg}(s), y_{seg}(s)]$ 
along a loop segment with constant curvature radius and length $n_s$. 
If the segment follows an ideal ridge with a constant curvature radius 
and a constant brightness, the summed (or averaged) flux along the ridge 
segment has a maximum value, while it exhibits a minimum in perpendicular 
direction to the ridge, where the brightness profile collapes to a 
$\delta$-function.}

\item{\underbar{Loop subtraction in residual image:} Once a full loop
structure has been traced, the loop area $I_0(x_i\pm w, y_j\pm w)$, 
$i=1,...,n_s$, is set to zero within a half width of $w=(n_{sm2}/2-1)$,  
so that the area of a former detected loop is not used in the detection 
of subsequent loops. However, crossing loops can still be connected over 
a gap.}

\end{enumerate}

The original automated loop tracing code (OCCULT) employed the following
control parameters: the highpass
filter boxcar $n_{sm2}$, the noise threshold level $N_\sigma$,
the minimum curvature radius $r_{min}$, the moving segment length $n_s$, 
the directional angle range $\Delta \alpha$, and a filling factor $q_{fill}$
for the guiding criteron. In the advanced code (OCCULT-2) we use only
two free parameters: the lowpass filter $n_{sm1}$, and the minimum 
curvature radius $r_{min}$. In addition, noise treatment is handled
by selecting a typical noise area in the image, as well as the control
factor $q_{med}$ that ignores faint structures below the base level 
$z_{min} = z_{med} \times q_{med}$. Thus, the new version of the code
offers a simpler choice of control parameters for automated
detection of structures. 

\section{Application to Solar Physics}

In the following three subsections we process three different types
of solar images, one from the TRACE spacecraft (Section 3.1), one
from the SDO spacecraft (Section 3.2), and one from a ground-based
solar telescope (Section 3.3). Some parameters of the analyzed images
are listed in Table 1. 

\begin{table}[h]
\caption{Parameters of five analyzed images, including the range of
brightness in the image ($z_{min},z_{max}$), the minimum curvature radius
$r_{min}$ (in units of pixels), the bandpass filter ($n_{sm1}, n_{sm2}$), 
the total number of detected loops $N_{det}(L>30)$, the number of detected 
long loops $N_{det}(L>70)$, and the powerlaw slope of the cumulative length 
distribution $p_L$.}
\begin{tabular}{lrrrrrr}
\hline
Image & Brightness & Minimum & Filter & Number of & 
Number of & Powerlaw\\
& range      & curvature & range  & all loops & long loops & slope\\
& $z_{min},z_{max}$ & $r_{min}$ & $n_{sm1},n_{sm2}$ & 
	$N_{det}(L>30)$ & $N_{det}(L>70)$ & $p_L$ \\
\hline
TRACE   &  56,  2606  &   30	&  5,  7  &  437   & 134   & 3.1   \\
SDO/AIA &  28,   255  &   30	&  9, 11  &  503   & 121   & 2.7   \\
SST	& 339,   916  &   30	&  3,  5  & 1757   & 376   & 3.9   \\
Cell-HC & 416, 65535  &   15    &  3,  5  &  208   &  51   & 3.1   \\
Cell-LC & 289, 18699  &   30    &  7,  9  &  151   &  39   & 2.7   \\
\hline
\end{tabular}
\end{table}

\subsection{TRACE Data}

The first image we are processing has been used as a standard in a
number of previous publications \cite{lee2004, lee2006, aschwanden2008,
aschwanden2010}, which shows coronal loops in a dipolar active region
that was recorded by the {\sl TRACE} spacecraft on 1998 May 5, 22:21 UT,
in the 171 \ang\ wavelength (Fig.~1, center). 
The original image has a size of
$1024 \times 1024$ pixels with a pixelsize of 0.5\arcsec ($\approx
360$ km on the solar surface). The EUV brightness or intensity in
every pixel is quantified by a data number (DN), which
has a minimum of 56 DN and a maximum of 2606 DN in this image.
We show the bandpass-filtered image
(with a lowpass boxcar of $n_{sm1}=5$ pixels and a highpass boxcar
of $n_{sm2}=7$ pixels) in Fig.~2. The image shows at least four different
textures (Fig.~1, side panels): coronal loops (L: curvi-linear features), 
so-called moss regions
(M: high-constrast reticulated or spongy features in the center of the
image, which represent the footpoints of hot coronal loops), 
transition region emission (T: low-contrast irregular features in the 
background), and faint emission areas.
The faint image areas show in addition a ripple with diagonal stripes
at a pedestal level of $\approx 57\pm1$ DN (Fig.~2 bottom left panel R)
that results from some interference in the electronic readout, which 
can produce unwanted non-solar structures in the automated detection 
of curvi-linear features. 

\begin{figure}
\centerline{\includegraphics[width=1.0\textwidth]{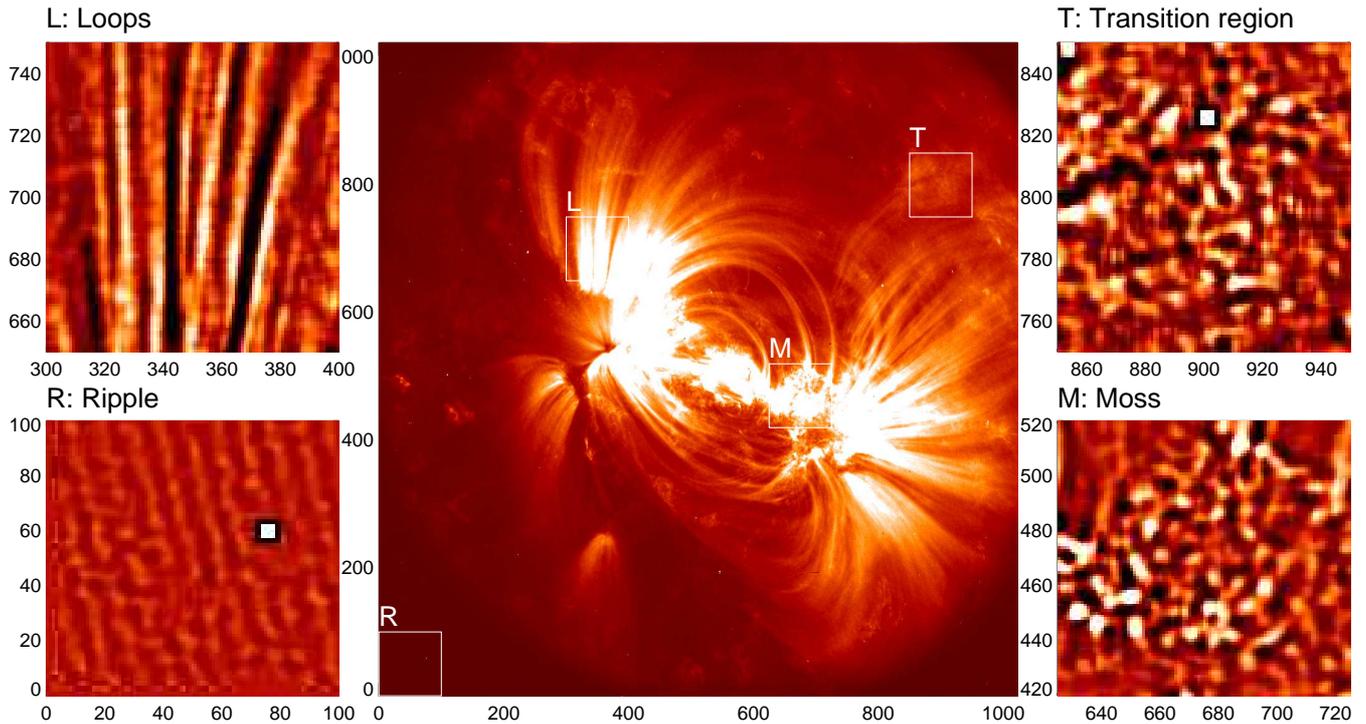}}
\caption{A solar EUV image of an active region, recorded with the 
TRACE spacecraft on 1998 May 15, is shown with a colorscale that has
the highest brightness in the white regions (center).
In addition we show ($100 \times 100$ pixel) enlargements of four
subregions with different textures, which contain coronal loops 
(top left panel), electronic ripple (bottom left panel), 
chromospheric and transition region emission (top right panel), 
and moss regions with footpoints of hot coronal loops 
(bottom right panel).}
\end{figure}

\begin{figure}
\centerline{\includegraphics[width=1.0\textwidth]{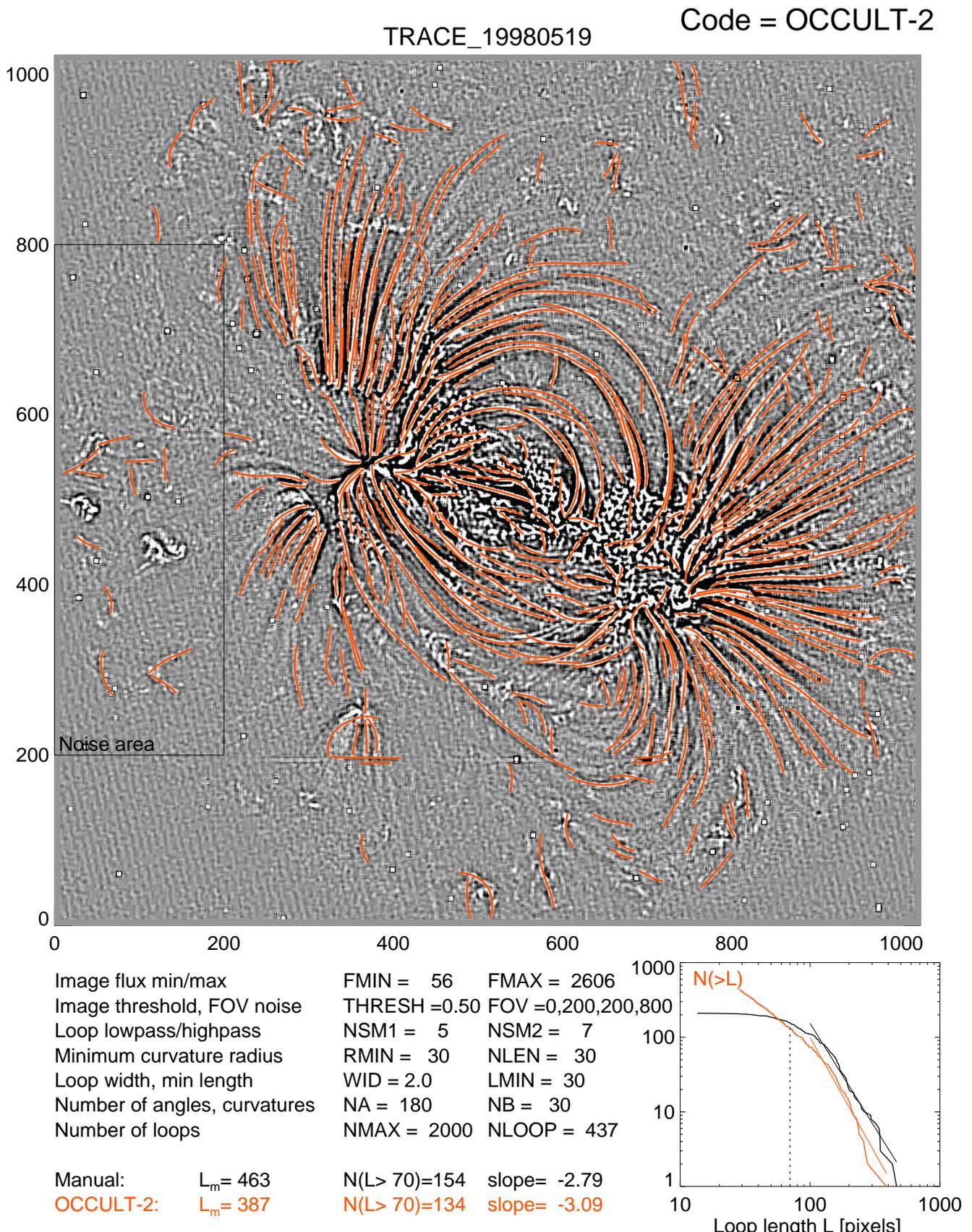}}
\caption{A bandpass-filtered ($n_{sm1}=5, n_{sm2}=7$) version of the
original image rendered in Fig.~1 is shown (greyscale),
with automated loop tracings overlaid (red curves). Cumulative
size distributions $N(>L)$ of loop lengths are also shown (bottom
right panel), comparing the automated tracing (red distribution) with
visually/manually traced loops (black distribution). The maximum
lengths $L_m$ (in pixels) are listed for the longest loops detected 
with each method.} 
\end{figure}

The automated detection of curvi-linear features with the OCCULT-2
code yields a total of 437 loop structures with lengths of 
$L \ge 30$ pixels (Fig.~2), whereof the longer loops (with lengths 
of $L \gapprox 50$) coincide well with the 210 visually/manually traced
loops (see Fig.7 in \cite{aschwanden2010}). The good agreement
between the automated and visually detected loops can also be seen
from the cumulative size distributions of loop lengths obtained with
both methods (Fig.~1, bottom right panel). The two methods detect $N=154$
and $N=134$ loops with a length of $L \ge 70$ pixels, and both distributions
have a powerlaw slope of $\alpha_L \approx 2.9\pm0.1$. 
Challenges of coronal loop detection in this image are confusion
and interference from all three types of non-loop structures (Fig.~1):
chains of ''dotted moss structures'', filamentary and spicular transition
region emission, and the diagonal stripes of electronic ripple 
in the background, which become all comparable with the
signal of loop structures once the image contrast is enhanced with
a highpass filter. 

\begin{figure}
\centerline{\includegraphics[width=0.8\textwidth]{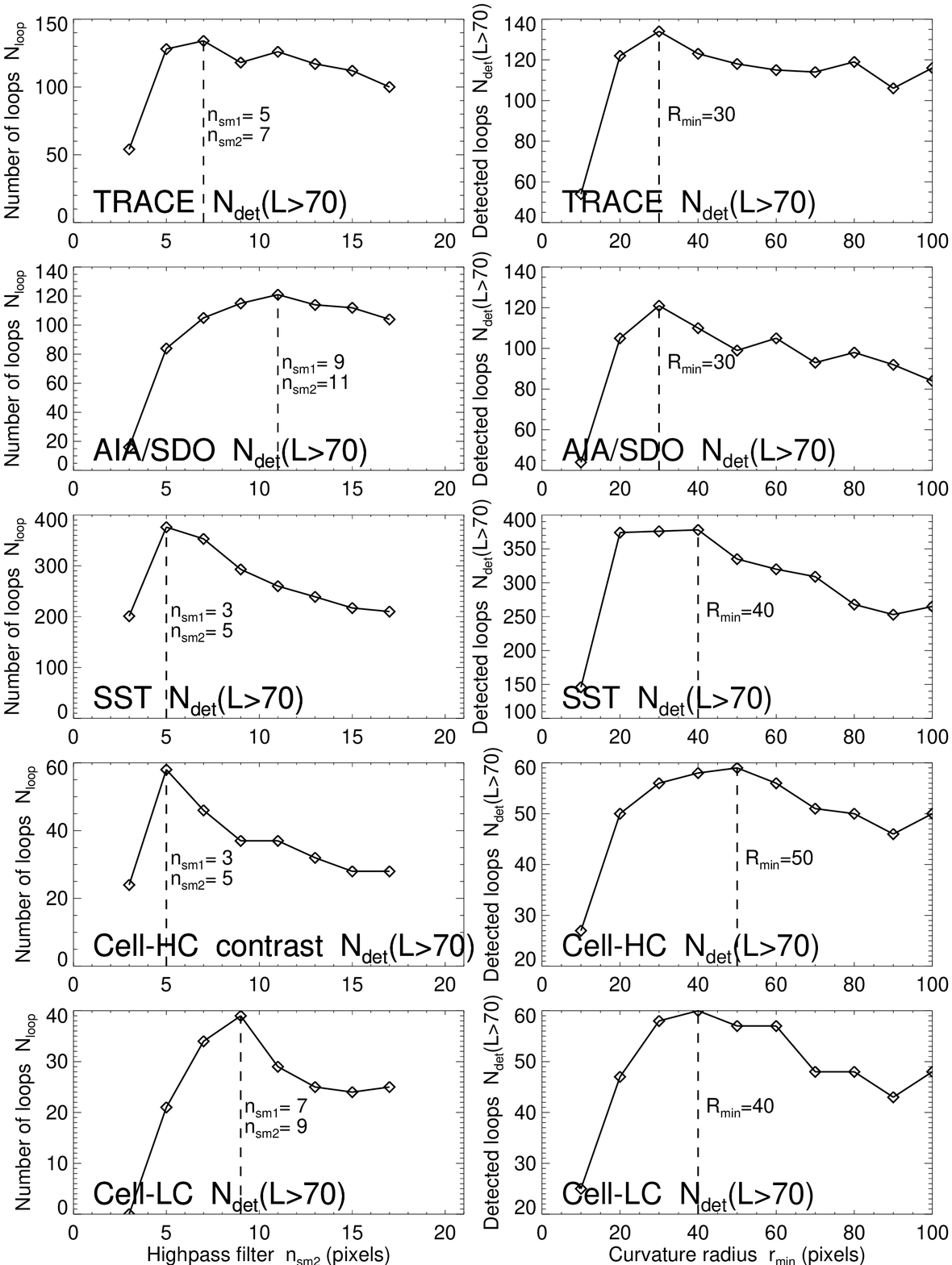}}
\caption{The optimization of the highpass filter constant $n_{sm2}$ 
(y-axis) is shown for the number of detected loops (with lengths longer 
than 70 pixels) for all analyzed cases, $N_{det}(L \ge 70)$  
(left panels). The optimization of detected loops as a function
of the curvature radius $r_{min}$ is also shown (right panels).}
\end{figure}

The successful or false detection in an image thus
depends most strongly on the chosen low and highpass filter constant
$n_{sm2}$ and the background base level 
$z_{min} = z_{med} \times q_{med}$, and to a lesser degree on the other
control parameters. A good indicator
of the completeness and efficiency of automated loop detection 
is the number of coherently detected long loops, say with a length 
above 70 pixels here, $N_{det}(L>70)$ (which is marked with a dotted line
in the size distribution in Fig.~2, bottom right panel). 
In Fig.~3 (top left panel)
we show how this detection efficiency $N_{det}(n_{sm2})$
varies as a function of the chosen control parameter  
$n_{sm2}$. The detected number has a
maximum of $N_{det}=134$ at $n_{smi1}=5$ and $n_{sm2}=7$, which
corresponds to a bandpass filter in the range of 5-7 pixels 
($2.5\arcsec-3.5\arcsec$ or 1800-2500 km on the solar surface).
It appears that this is the most typical cross-section width (FWHM) 
of coronal loops. Other statistical studies of coronal loops observed
with TRACE yield similar values (e.g., FWHM=$1420\pm340$ km; 
\cite{aschwanden2005}). Thus, if an image contains structures with a
preferential cross-section width, the relevant cross-section range
can be bracketed with a bandpass filter ($n_{sm1}, n_{sm2}$), 
providing a useful {\sl a priori} information for automated detection of
curvi-linear features. We conducted tests with all possible filter widths
$n_{sm1}=1,3,...,21$ and $n_{sm2} > n_{sm1}$ and found that the largest
number of detected structures virtually always occurs at 
$n_{sm2} = n_{sm1}+2$, which can be explained also by the theoretical
argument that the best signal-to-noise ratio is obtained for maximum
smoothing of the highpass-filtered (unsharp mask) image.
We vary also the minimum curvature radius $r_{min}$ and find a maximum 
of detected structures at $r_{min} \approx 30$ pixels (Fig.~3 top right 
panel). 

\begin{figure}
\centerline{\includegraphics[width=1.\textwidth]{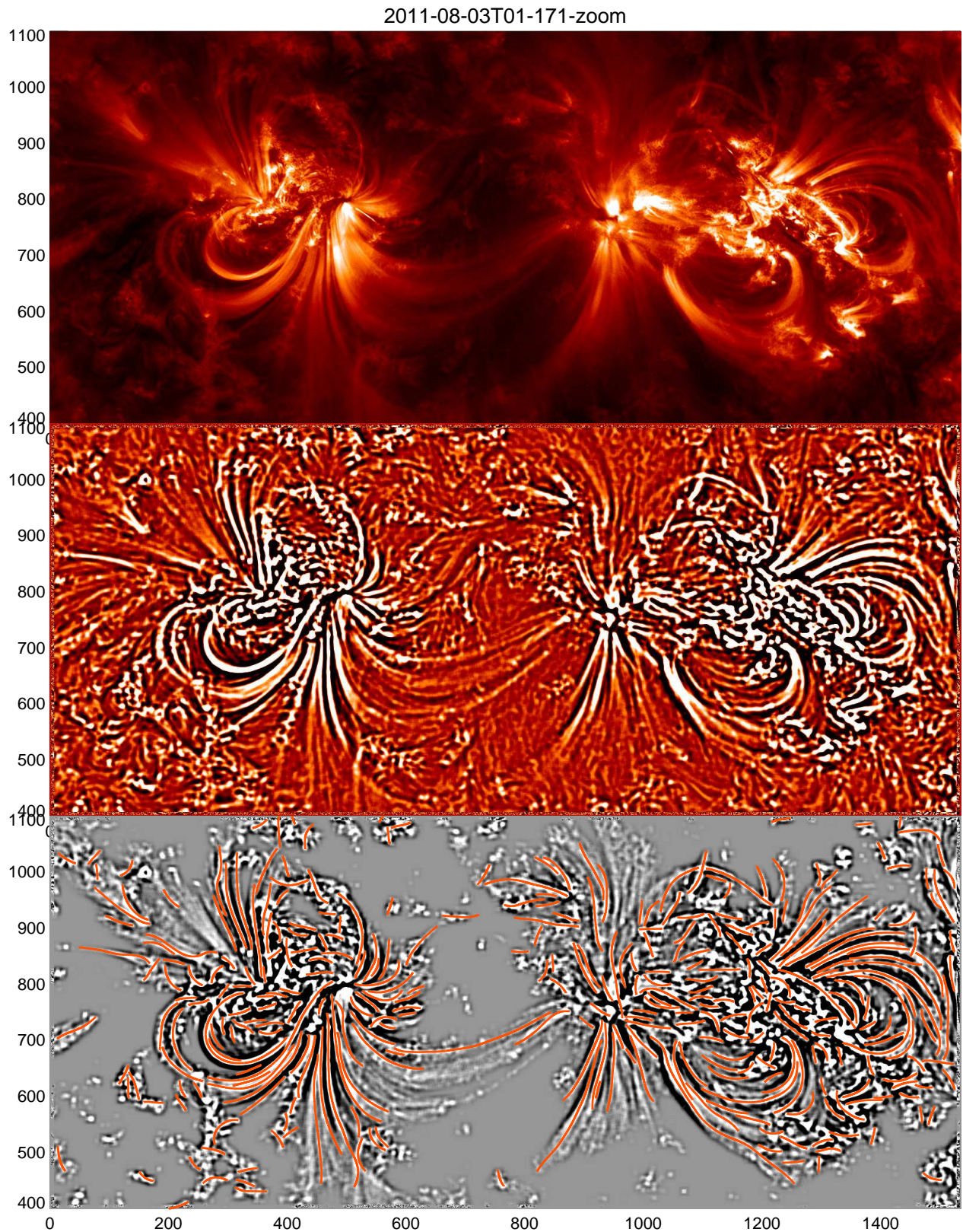}}
\caption{Bandpass-filtered image of an active region complex observed
with AIA/SDO on 2011 Aug 3, 01 UT, 171 \ang\ , shown as intensity image
(top panel), as bandpass-filtered version with $n_{sm1}=9$ and
$n_{sm2}=11$ (middle panel), and overlaid with automatically traced
loop structures (bottom panel), where the low-intensity values below 
the median of $f=75$ DN are blocked out (grey areas).}
\end{figure}

\subsection{SDO/AIA Data}

The next image to which we apply our automated loop tracing code is
from the {\sl Atmospheric Imager Assembly (AIA)} onboard the {\sl Solar
Dynamics Observatory (SDO)}, which replaced the TRACE mission and is
operating since 2010 \cite{lemen2012}. AIA has a similar spatial
resolution (pixel size 0.6\arcsec) as TRACE (pixel size 0.5\arcsec),
but covers the full Sun disk, with an image size of $4096\times 4096$
pixels. Fig.~4 shows a subimage with a size of $1450 \times 650$ pixels,
which contains a complex of two magnetically coupled active regions,
observed on 2011 Aug 03, 01 UT, in the 171 \ang\ wavelength.
This image is currently subject of nonlinear force-free magnetic modeling
(Mark DeRosa, private communication 2012), and thus requires automated
loop tracing to constrain the coronal part of the magnetic field
configuration \cite{aschwanden2013a}. Differences to the TRACE image 
are the higher sensitivity of the AIA telescopes, different exposure 
times, the availability of simultaneous images in 8 other wavelengths,
different image compression, and no apparent electronic ripple in the 
CCD readout, which all affect the automated detection of faint structures.
Synthesized loop tracings from multiple wavelength filters has been proven
to provide a more robust and representative subset of loop structures 
for magnetic modeling than loop tracings from a single filter image
(\cite{aschwanden2013b}).

We vary the lowpass filter constant in the range of $n_{sm1}=1,...,21$,
set the highpass filter constant to $n_{sm2}=n_{sm1}+2$,
and find a maximum detection rate of $N_{det}(L>70)=121$ loop structures
for $n_{sm1}=9$ and $n_{sm2}=11$ (Fig.~3, second row left panel). 
We vary also the minimum curvature radius in the
range of $r_{min}=10,...,100$ pixels and find a maximum detection rate
of $N_{det}(L>70)=121$ at $r_{min}=30$ pixels (Fig.~3, second row right
panel). 

\begin{figure}
\centerline{\includegraphics[width=1.\textwidth]{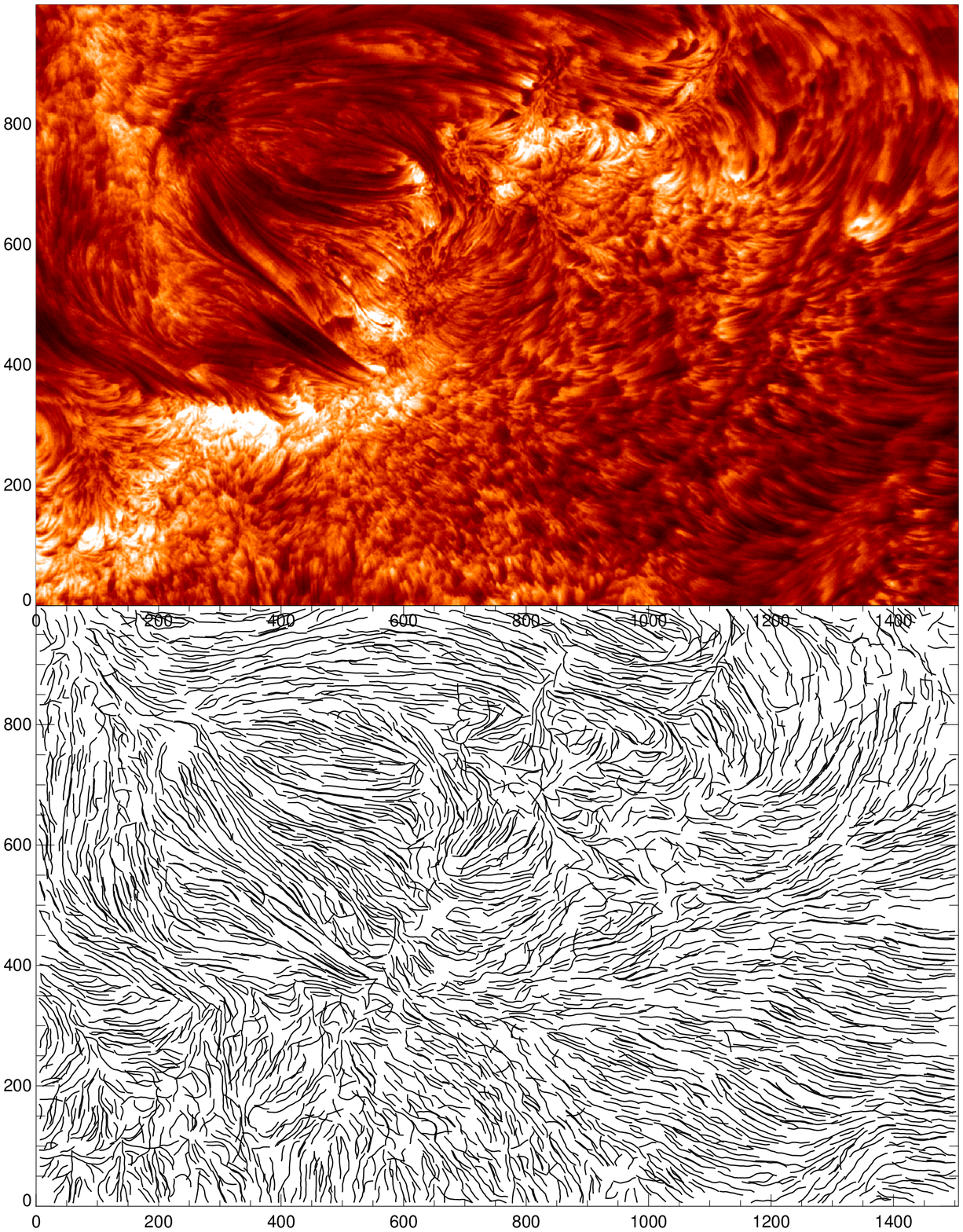}}
\caption{High-resolution image of the solar Active Region 10380, 
recorded on 2003 June 16 with the Swedish 1-m Solar Telescope (SST)
on La Palma Spain (top panel) and automated tracing of curvi-linear
structures with a lowpass filter of $n_{sm1}=3$ pixels, a highpass
filter of $n_{sm2}=5$ pixels, and a minimum curvature radius of
$r_{min}=30$ pixels, tracing out 1757 curvi-linear segments (bottom panel).}
\end{figure}

\subsection{SST Data}

Now we apply automated loop tracing to a solar image in a completely
different wavelength, namely in the H$\alpha$ 6563 \ang\ line, 
the first line of the Balmer series of hydrogen. Fig.~5
shows such an image of the solar upper chromosphere, which displays
chromospheric spciules in the right side of the picture and 
filamentary structures in the upper chromosphere and transition region
(in altitudes $\approx 2000-5000$ km above the solar surface) 
\cite{depontieu2004, depontieu2006}. 
The image was taken with the {\sl Swedish 1-m Solar
Telescope (SST)} on La Palma, Spain, using a tunable filter, tuned
to the blue-shifted line wing of the H$\alpha$ 6563 \ang\ line. The
spicules are jets of moving gas at a lower temperature than the
million degree hot corona and flow upward from the chromosphere
to the transition region with a speed of $\approx 15$ km s$^{-1}$
\cite{depontieu2004}.
The image has a size of $1507 \times 999$ pixels ($62 \times 41$
Mm on the solar surface), with a pixel size of $0.041\arcsec$ (or 30 km
on the solar surface). This picture is particularly intriguing for
automated tracing of curvi-linear structures because of the ubiquity
and complexity of fine structure.

This SST image has such a high contrast so that there is no significant
noise that affects curvi-linear tracing. Thus, we set the 
background level to zero ($q_{med}=0.0$) in the OCCULT-2 code.
We vary the lowpass filter constant in the range of $n_{sm1}=1,...,21$,
set the highpass filter constant to $n_{sm2}=n_{sm1}+2$,
and vary the minimum curvature radius in the range of $r_{min}=10,...,100$
pixels. We find a maximum number of detected structures of
$N_{det}(L>70)=376$ (with a length above 70 pixels) at
$n_{sm1}=3$ and $N_{sm2}=5$ (Fig.~3, middle row left panel), and
$r_{min}=40$ pixels (Fig.~3, middle row right panel). 
Extending to shorter segments with lengths of $L>30$ pixels, the 
automated tracing code identifies a total of 1757 curvi-linear segments,
which outline the patterns of the flow field in the upper chromosphere
(Fig.~5, bottom panel).

\begin{figure}
\centerline{\includegraphics[width=1.\textwidth]{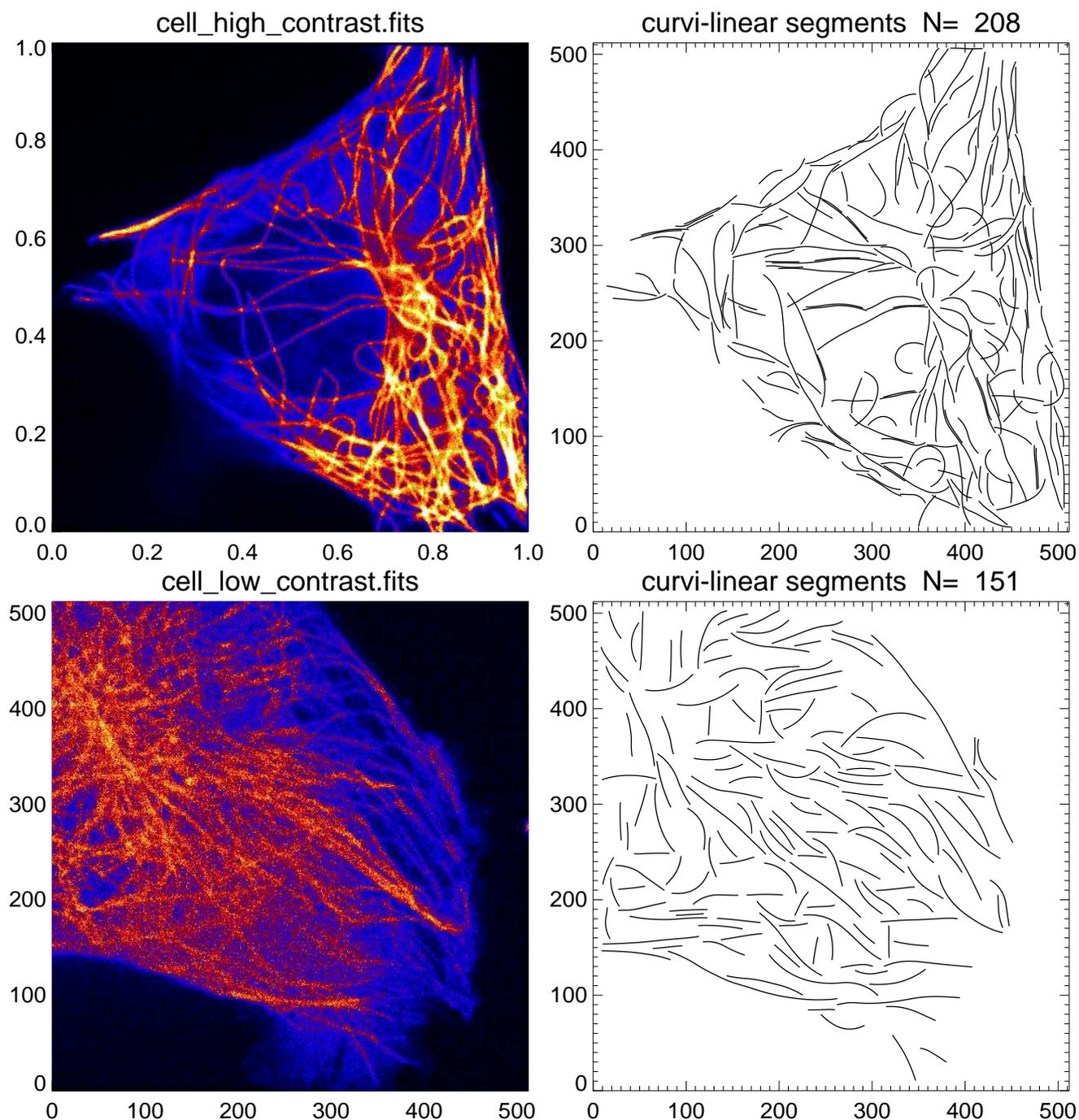}}
\caption{False-coloured images of microtubule filaments in
live cells: a high-contrast image of CHO cell (top left panel, Cell-HC), 
a low-contrast image of U2OS cell (bottom left panel; Cell-LC). The size of 
the images corresponds to 34 $\mu$m. The automated curvi-linear 
tracing of both images was carried out with the parameters: 
$n_{sm1}=3$, $n_{sm2}=5$, and $r_{min}=15$ for Cell-HC, and 
$n_{sm1}=7$, $n_{sm2}=9$, and $r_{min}=30$ for Cell-LC.}
\end{figure}

\section{Applications to Biophysics}

Finally we apply our automated tracing code to images obtained in 
cellular biophysics, in order to test the versatility and universality 
of the OCCULT-2 code. As an example, we chose microscopy images of 
microtubules, but the same approach could be applied to any other 
filament-like structures: intermediate or actin filaments, fibrin, etc. 
Microtubules are long stiff polymers that are part of cytoskeleton 
(internal cellular scaffolding). They are important for cell division, 
motility and organization (\cite{kaverina2011}, \cite{deforges2012}). 
The entangled network of microtubules serves as a ``railroad'' system 
for delivery of cargos by molecular motors and also as a stiff carcass 
controlling cellular mechanics (\cite{deforges2012}, 
\cite{subramanian2012}). It is a dynamic network adapting and changing 
in time in response to external cues. The automated extraction of the 
microtubules network's configuration from microscopy images and movies 
can provide insights on the mechanisms of these changes. 

Fig.~6 shows two images of cells from two different cell lines with 
microtubules networks of different density. Cells belonging to CHO cell 
line (Fig.~6 top, image Cell-HC) are small and have sparse microtubule 
network. Cells from U2OS line (Fig.~6 bottom, image Cell-LC) are larger 
and contain more dense radial network. Since microtubules are transparent 
to the visible light, a fluorescent tag is used to observe them in the 
living cells. In the analyzed images (Fig.~6) microtubules were labeled 
with a green fluorescent protein (GFP) (\cite{tsien1998}) that adsorbs 
light at 490 nm and has an emission peak at 510 nm wavelength. The images 
were acquired with a spinning disk confocal microscope using the 
corresponding GFP's emission filter. The magnified image was projected 
on the 16-bit chip of camera with dimensions of $512 \times 512$ pixels 
resulting in the final image pixel size of 66 nm. The width of 
microtubules filaments measured in the electron microscopy studies is 
about 25 nm (\cite{wade1993}). Due to the diffraction limit the effective 
width of microtubules in the image is much larger and is defined by the 
microscopy setup and the wavelength used (\cite{abbe1884}). In our case 
it is approximately equal to a half of the emission wavelength (510 nm) 
that is close to the measured microtubule width of $4.7 \pm 2.0$ pixels. 
The two pictures shown in Fig.~6 (left) show cases of opposite contrast, 
one with high contrast (Fig.~6 top, image Cell-HC), and one with low 
contrast (Fig.~6 bottom, image Cell-LC). The difference in contrast is 
explained by a different amount of GFP fluorescent tag associated with 
microtubules. The automated tracing of the cell filaments provides 
their locations inside cell and curvature radii, from which flexural 
rigidity and mechanical stress can be inferred.

We optimize the automated filament tracing by varying the bipass filter
constants and find a maximum number of detected filaments (with lengths
$L>70$ pixels) with $n_{sm1}=3$, $n_{sm2}=5$, $r_{min}=50$ for the
high-contrast cell image, and $n_{sm1}=7$, $n_{sm2}=9$, $r_{min}=40$ 
for the low-contrast cell image (Fig.~3). The low-contrast image has
a higher degree of noise, and thus the filter constants have to be
adjusted to a larger value for the low-contrast cell image ($n_{sm1}=7$)
than for the high-contrast image ($n_{sm1}=3$). 

\section{Discussion and Conclusions}

\subsection{Optimization of Automated Curvi-Linear Tracing}

The efficiency and accuracy of automated curvi-linear tracing can be
controlled by a number of tuning or control parameters. For the
present OCCULT-2 code we have three independent control parameters
($n_{sm1}, r_{min}, q_{med}$).
A prerequisite for the application of curvi-linear tracing codes
is the assumption that the structures of interest have a much
smaller width $w$ than their length $l$, so that they can be
represented by a 1D path, which is generally curved, possibly
limited by a minimum curvature radius $r_{min}$. The larger the
minimum curvature radius is, the less ambiguity there is for tracing
of crossing 1D structures. In this study we explored the
parameter space of the control parameters $n_{sm1}$ and $r_{min}$ 
in order to optimize the performance of the automated tracing code. 

The lowpass filter ($n_{sm1}$) and highpass filter ($n_{sm2}$)
represent the brackets or scale range of a bandpass filter, bracketing
the range of cross-sections of detected structures. We expect to find
structures with the smallest width preferentially in images with a 
high signal-to-noise ratio, while noisy images require more smoothing
to enhance fine structure, and thus tend to have larger widths due to
the smearing effect of the smoothing.
We found the narrowest structures indeed in the two images
with the highest contrast (i.e., the SST and Cell-HC image), where
highpass filters with boxcars of $n_{sm2}=5$ pixels were used. In images 
with lower contrast, optimum performance occurred for highpass 
filters of $n_{sm1}=7,...,11$ pixels. In addition, we found that the
smallest bandpass filters produce the sharpest structures and thus
the highest detection rate of structures. Since a symmetric boxcar
requires odd numbers $n_{sm}=1,3,5,...$, the smallest difference between
a lowpass and a highpass filter is 2, and thus the optimum combination
is expected to be $n_{sm2}=n_{sm1}+2$, which we indeed confirmed also
experimentlly. 

For the minimum curvature radius we found optimum performance
for a typical range of $r_{min} \approx 30,..,50$ pixels (Fig.~3),
which seems not to depend on the contrast of the image.

The control parameter $q_{med}=1$ suppresses faint structures in
an image that are below the median value of the image brightness.
A meaningful value for this parameter depends very much what
fraction of the image contains bright structures of interest,
which has to be decided depending on the area ratio of structures
of interest to non-relevant background area. If the background
has comparable brightness to the structures of interest, a
thresholded separation may be impossible, but may still succeed
if the background has a different texture than the curvi-linear
features (see examples in Fig.~1, where moss and transition region
emission have a different texture than coronal loops, while background
ripples have the same texture as straight loops and can only be
separated by a threshold control parameter). Future efforts aim
to synthesize the loop tracings from  multi-wavelength image sets, 
which are more robust and representative for magnetic modeling
than single-wavelength images (\cite{aschwanden2013b}).

In conclusion, we recommend the following procedure to achieve
optimum performance of the curvi-linear tracing code (OCCULT-2):
(1) Start with the following recommended control settings:
$n_{sm1}=1$, $n_{sm2}=3$, $r_{min}=30$, and $q_{med}=1.0$; 
(2) Vary the filter combination in the range of $n_{sm1}=1,3,...,15$,
while setting $n_{sm2}=n{sm1}+2$, to find the maximum detection rate 
for a given loop length (e.g., here we used $L>70$ pixels);
(3) Low-contrast images are likely to require higher highpass
filter values $n_{sm2}$ than high-contrast images.
(4) Vary the minimum curvature radius $r_{min}$ within some range
to find the maximum detection rate of structures;
(5) If the code yields a lot of random structures in obvious 
background areas, increase the base level factor $q_{med} > 1$. 

The software of the numerical code OCCULT-2 is publicly accessible
in the {\sl Interactive Data Language (IDL)} in the {\sl Solar
Software (SSW)} package. A tutorial and example is accessible
at the authors homepage {\url http://lmsal.com/$\sim$aschwand/software/}.

\subsection{Solar Applications}

The automated tracing of curvi-linear structures in solar physics
has mostly been applied to extreme-ultraviolet or soft X-ray images,
which show magnetized coronal loops that outline the coronal magnetic
field, due to the low plasma-$\beta$ parameter in the solar corona
(which is defined as the ratio of the thermal to the magnetic pressure).
Thus, coronal loops represent the perfect tracers of the otherwise
invisible coronal magnetic field. Standard magnetic field models of
the solar corona or parts of it, such as sunspot regions and active
regions, have been modeled by extrapolating a photospheric magnetogram,
either with a potential field solution or a force-free solution
of Maxwell's equations. However, since the chromosphere was found
not to satisfy the force-free condition \cite{metcalf1995}, 
extrapolations of photospheric magnetograms do not exactly render
the coronal magnetic field, while coronal loops outline the true
coronal magnetic field. It is therefore desirable to trace such
coronal loops in EUV and soft X-ray images and to use them to constrain
a magnetic field solution. Such attempts have been performed with
single EUV images as well as with stereoscopic EUV image pairs
\cite{aschwanden2013a}. Forward-fitting of theoretical magnetic field
models to traced coronal loops is able to discriminate between
potential field and force-free field models, as well as to quantify
the free magnetic energy (that is released in solar flares)
and Lorentz forces, e.g., \cite{gary2009}.
Curvi-linear tracing of filamentary structures
in the chromosphere and transition region (Fig.~5) may also help to
constrain the horizontal magnetic field components in the non-forcefree
regions, which has been used in pre-processing of solar force-free
magnetic field extrapolations \cite{fuhrmann2011}.

One fundamental limitation of automated coronal loop tracing is
the confusion by background structures resulting from EUV emission
from the transition region, which has generally a cooler temperature
than the coronal loops. Future efforts may use multi-wavelength 
image data sets to discriminate EUV emission from the chromosphere,
transition region, and the corona by its temperature, using a
a deconvolution of the multi-wavelength temperature filter 
response functions in terms of a differential emission measure 
(DEM) method.

\subsection{Biological Applications}

The method of curvi-linear tracing is increasingly used in the 
analysis of biological and medical images, such as to characterize
blood vessel tracking in retinal images \cite{jiang2007, tang2004,
martinez1999}, neurons \cite{xiong2006}, dendritic spines \cite{zhang2007},
or microtubule tracing in fluorescent and phase-contrast microscopy 
\cite{brangwynne2007, sargin2007, koulgi2010}, as shown in Fig.~6. 
In our experiment
with a high-contrast (Fig.~6, top panel) and low-contrast image (Fig.~6,
bottom panel), we demonstrated that a bipass or bandpass filter 
with a bandpass factor of $n_{sm2}=n_{sm1}+2$ enhances the
structures to an optimum contrast for automated tracing. 
Moreover we found that the bandpass filter
of low-contrast images requires a larger width ($n_{sm2}=9$ pixels
in Fig.~6, bottom panel) than in a high-contrast image ($n_{sm2}=5$ pixels
in Fig.~6, top panel). 

In our optimization exercise we concentrated
mostly on the completeness of detected long curvi-linear segments,
but future efforts may also consider the optimization of
linking multiple segments that are interrupted with gaps or subject to
crossings. The efficiency and reliability of automated curvi-linear 
algorithms became more important with the massive increase of
imaging data over the last decade, which exceeds our limited 
capabilities of visual inspection. Note that the algorithm used here
tracks curvi-linear structures that differ by 16 orders of magnitude
in size, from 66 nm to 360 km (pixel size in images).

\section{Appendix A: Analytical Description of the OCCULT-2 Code}

The {\sl Oriented Coronal CUrved Loop Tracing (OCCULT-2)} code version 2
is an improved
version of the original OCCULT code described in Aschwanden (2010). The
improvements include: (1) a ``curved'' guiding segment that is adjusted to the 
local curvature radius of a traced loop (representing the second-order
term of a polynomial), rather than the linear (first-order polynomial)
guiding segment used in the original version, (2) suppression of faint 
structures, (3) bypass filtering instead of highpass filtering, and 
(4) simplification of selectable free parameters.
  
The input is a simple 2-dimensional (2D) image $z_{ij}$,
with pixel numbers $i=0,...,n_x-1$ on the x-axis, and $j=0,...,n_y-1$ on 
the y-axis, respectively. The output is a number of curvi-linear 
structures (also called ``loops'' for short), which are parameterized 
in terms of x and y-coordinates, $[x(s_k),y(s_k)]$, where the loop length 
coordinate $s_k=0,...,n_s$ is given in steps of $\Delta s=1$ pixel, 
so that for all $k=0,...,n_s-1$, 
\begin{equation}
	\Delta s_k =\sqrt{\left( [x(s_{k}-x(s_{k-1})]^2 
                               + [y(s_k)-y(s_{k-1})]^2 \right) } = 1,
	\quad k=0,...,n_s-1 \ ,
\end{equation}
with $n_s$ the number of loop points for each loop.  

The goal of the algorithm OCCULT-2 is to retrieve as many curvi-linear 
structures as possible, without picking up false signals of non-existing 
structures in the noise of the image, which has some probability to form 
chains of random points in a curved array configuration. The challenges 
are therefore to evaluate an optimum threshold level that separates 
existing loops from noise structures, and to retrieve the 
real curvi-linear structures as completely as possible, without subdividing
them into partial loop segments. Our strategy to obtain a fast numeric code 
is to retrieve the loops in a one-dimensional search algorithm, because 
any two-dimensional concept has a computation time that grows with the 
square of the image size. The one-dimensional parameter space is essentially 
the loop length coordinate $s_k$, $k=0,...,n_s-1$. In addition we define 
two other independent parameters in each loop point, which can 
be considered as the first-order and second-order term of a polynomial,
namely the local direction angle $\alpha_l$, 
$l=0,...,n_\alpha$, and the curvature radius $r_m$, $m=0,...,n_r$. 
The algorithm selects iteratively the brightest position $(x_i, y_j)$ 
in the image and starts a bi-directional loop tracing, determining first 
the local direction $\alpha_l(x_i,y_j)$ and curvature radius $r_m(x_i,y_j)$ 
at the starting point, and continues tracing the loop within a small 
(guided) range of the local curvature radius, which is the principle of 
``orientation-guided tracing''. So we are dealing essentially with 1D-tracing 
in a 5D-parameter space $(x_i,y_j,s_k,\alpha_l,r_m)$, which we parameterize 
with the 5 indices $(i,j,k,l,m)$ that have the index ranges $i=0,...,n_x-1, 
j=0,...,n_y-1, k=0,...,n_s, l=0,...,n_\alpha, m=0,...,n_r$. Specifically, 
we define the arrays,
\begin{equation}
	s_k^{bi} = \Delta s \ (k - {n_s \over 2}) \ ,
	\quad k=0,...,n_s-1 \ ,
\end{equation}
for a symmetric bi-directional array, used in the search of the local 
direction $\alpha_l$, and 
\begin{equation}
	s_k^{uni} = \Delta s \ k \ ,
	\quad k=0,...,n_s-1 \ ,
\end{equation}
for a uni-directional array, used in the search for the curvature radius in 
the forward-direction of a traced structure. For the directional angle 
$\alpha_l$, which is only determined at the starting point of the loop, 
we use a fixed array with a resolution of one degree (or $\pi/180$ radian),
\begin{equation}
	\alpha_l = \pi \left( {l \over n_\alpha} \right) \ , 
	\qquad l=0,...,n_\alpha \ , \quad n_\alpha=180 \ .
\end{equation}
For the curvature radii $r_m$ we use a reciprocal scaling in order to obtain 
a uniform distribution of directional angles at the end of a curvature segment,
\begin{equation}
	r_m = {r_{min} \over [ - 1 + 2 m/(n_r-1) ]} \  , \qquad n_r=30 \ .
\end{equation}
This parameterization covers positive and negative curvature radii in the 
ranges of $[-\infty, -r_{min}]$ and $[r_{min}, +\infty]$. The choice of 
an even number $n_r$ prevents the singularity $r_{m}=\pm \infty$.

Now we describe the consecutive steps of the algorithm one by one, which follow more or less the same flow chart as depicted in Figure 1 of Aschwanden (2010).

\medskip
\underbar{(1) Image base level $(q_{base}$):} If the image consists of 
high-contrast structures without unwanted secondary structures in the 
background, we do not have to worry about the image base level and can 
set it to the lowest value (which should be zero or positive in 
astrophysical images that record an intensity or brightness). However, 
if there are unwanted structures at a faint brightness level, we can 
just set the image base level $z_{base}$ above the brightness level 
of unwanted structures, which we parameterize with the factor $q_{med}$ 
in units of the median brightness level $z_{med}=median[z_{i,j}]$,
\begin{equation}
	z_{base} = z_{med} \times q_{base} \ ,
\end{equation}
so that the corrected brightness $z_{i,j}'$ in each pixel fulfills the 
condition
\begin{equation}
	z_{i,j}' \ge z_{base} \ .
\end{equation}
For $q_{med}=0$ the image is unchanged, while the image $z_{i,j}'$ appears 
to be flat in the fainter half area of the image, if set $q_{med}=1.0$. 
So, the value $q_{med}$ can be adjusted depending on the estimated 
fraction of the image that is covered with structures of interest. 
For solar images, this feature offers a convenient way to filter out 
coronal loops in active regions (which are bright) from unwanted 
structures in the Quiet Sun (which are faint).  

\underbar{(2) Bandpass Filtering ($n_{sm1}, n_{sm2}$):} The tracing of 
curvi-linear structures is considerably eased by enhancing of fine 
structures within a chosen bandpass that corresponds to the typical 
width $n_w$ of structures of interest, which is typically a few pixels, 
and assuming that the curvi-linear structures has a much longer length 
$n_s$ than width, i.e., $n_s \gg n_w$. We accomplish the enhancement 
with a bandpass filter, which consists of a lowpass filter with a boxcar 
$n_{sm1}$ and a highpass filter with a boxcar $n_{sm2}$,
\begin{equation}
	z_{i,j}^{filter}=smooth[z_{i,j}',n_{sm1}] - smooth[z_{i,j}', n_{sm2}] 
	\ ,
\end{equation}
which filters out broad structures with widths $n_w \gapprox n_{sm2}$ 
(highpass filter), but smoothes out fine structure with a boxcar of 
$n_{sm1} < n_{sm2}$.  We experimented with a large number of combinations 
$(n_{sm1}, n_{sm2})$ and found that the optimum choice is,
\begin{equation}
	n_{sm2} = n_{sm1} + 2 \ ,
\end{equation}
which follows the principle of maximum possible smoothing of fine structures 
with a given width $n_{sm2}$. The values for the smoothing with a symmetric 
boxcar has to be an odd integer, i.e., $n_{sm1}=1,3,5,...)$, which implies 
$n_{sm2}=3,5,7,...$, where the lowest value $n_{sm1}=1$ corresponds to the 
original image without any smoothing. This experimentally tested relationship 
for the optimum choice of bandpass filters $(n_{sm1}, n_{sm2})$ reduces also 
the possible parameter space by one dimension, and thus we have search only 
for $n_{sm1}=1,3,5,... $, while using $n_{sm2}=n_{sm1}+2$.

\underbar{(3) Noise Threshold:} Our algorithm starts with the brightest 
curvi-linear structure and proceeds to fainter structures, and thus we 
have to find a stop criterion that halts the procedure when it reaches 
the level of data noise. Such a noise threshold has to be determined 
empirically for every image, since there are many sources of possible 
data noise. Testing many images with completely different data types, 
we found that a most reasonable threshold level can be determined by 
an interactive choice of a noise area in the image that contains typical 
data noise but little structures of interest. Such an image area can 
be characterized by the pixel ranges $(i_{n1}:i_{n2}, j_{n1}:j_{n2})$. 
In this noise area we determine the median brightness $z_{med}^{noise}$ 
and define a noise threshold at the doubled value,
\begin{equation}
	z_{thresh} = 2 \times (z_{med}^{noise} > 0) \ ,
\end{equation}
using only the pixels with positive values in the noise area (in order 
to prevent a too low value of $z_{thresh}=0$ in the case when more 
than 50\% of the noisy pixels are below the previously chosen base value 
$z_{base}$). The rationale for the factor 2 in the threshold level 
comes from the fact that the median separates out only half of the 
noisy pixels, while the double value would separate out all noisy 
pixels if the distribution of noisy pixel values follows a linear 
relationship. 

\underbar{(4) Start of Curvi-Linear Structures:} We are ready now to 
trace the first curvi-linear structure. We determine the location 
$(i_0,j_0)$ of the absolute brightness maximum $z_0$ in the 
bandpass-filtered image $z_{i,j}^{filter}$,
\begin{equation}
	z_0 = z(x_0,y_0) = max[z_{i,j}^{filter}(x_i,y_j)] \ .
\end{equation} 
The rationale for the choice of this starting point is the expectation 
to trace first the most significant structure in the bandpass-filtered 
image, which can then be continued by going to the next-significant 
structure, once the tracing of the first structure has been successfully 
completed and the corresponding loop area is eliminated in a residual 
image. Consequently, the maximum of the brightness in the residual 
image marks the second-brightest structure and we can repeat the 
same procedure by tracing the next loop.

\underbar{(5) Loop Direction at Starting Point:} The next element of 
the structure to be traced is the first-order term of a polynomial, 
the direction angle $\alpha_l$, which is also the direction of a 
possible ridge that outlines the local segment of the structure. 
We determine this directional angle simply by measuring the flux 
averaged over a straight loop segment symmetrically placed over 
the starting point $(i_0, j_0)$ and rotated over a full range of 
possible angles $\alpha_l$ from $0^\circ$ to $180^\circ$ degrees. 
The x,y-coordinates of this linear segment are, with the array
$s_k^{bi}$ defined by Eq.~(5),
\begin{equation}
	x_{k,l} = i_0 + s_k^{bi} \cos{ \alpha_l} \ ,
\end{equation}
\begin{equation}
	y_{k,l} = j_0 + s_k^{bi} \sin{ \alpha_l} \ ,
\end{equation}
where the index $k$ runs along the length $s_k$ of the segment, and 
the index $l$ denotes a particular angle $\alpha_l$. Among the set of 
angular values $\alpha_l$ we determine the maximum of the summed flux 
in each rotated segment,
\begin{equation}
	z_{max}(\alpha_l) =max_l{\left[ {1 \over  n_s} \sum_{k=0}^{n_s-1} 
	z^{filter}(x_{k,l},y_{k,l}) \right]} \ ,
\end{equation}
which  yields the local direction $\alpha_{max}=\alpha_l(l=l_{max})$. 
In the ideal case of a straight ridge with a constant value $z_0$ along 
the ridge segment with length $n_s$ pixels and zero-values outside, 
a value of $z_\parallel=z_0$ is found, while the value for a segment in 
perpendicular direction to the ridge is much smaller, namely 
$z_\perp=1/n_s$. This ridge criterion works even for a close succession 
of parallel ridges, in which case it is still a factor of 2 smaller 
than in parallel direction, i.e., $z_\perp=1/2$. 

\begin{figure}
\centerline{\includegraphics[width=1.0\textwidth]{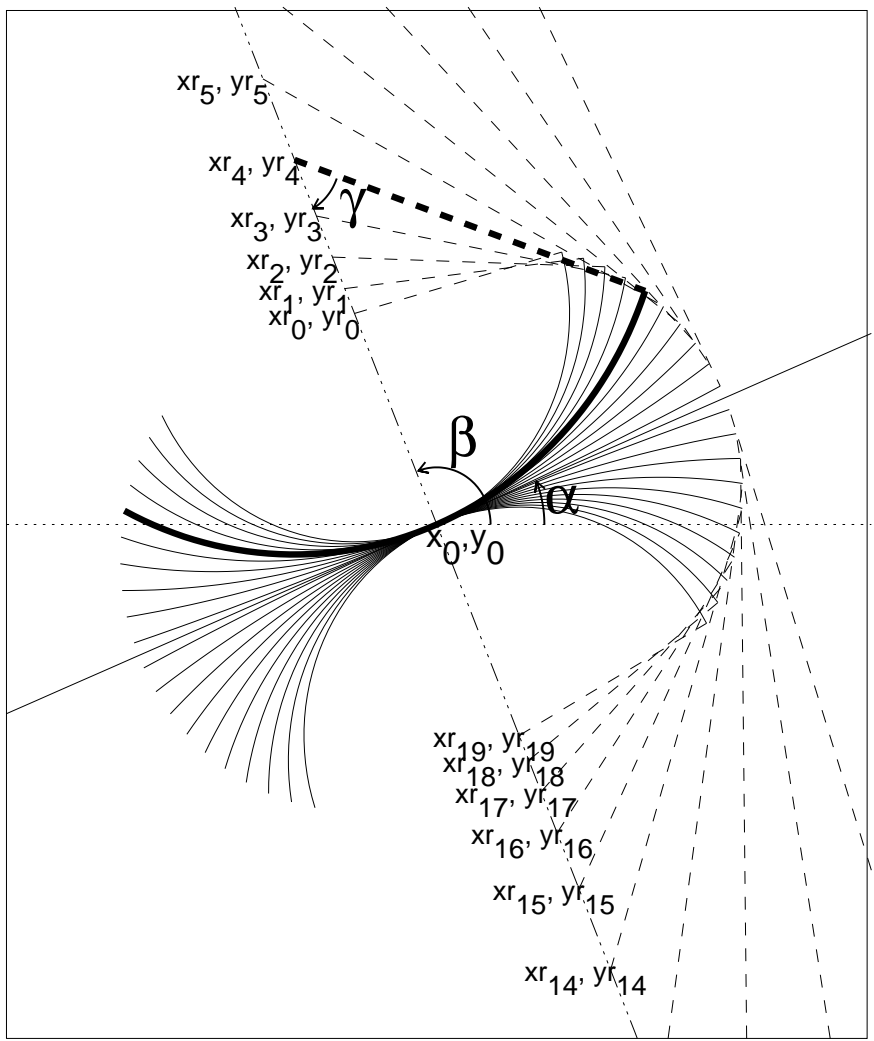}}
\caption{Geometry of curvature radii centers $(xr_m,yr_m)$ located
on a line at angle $\beta$ (dash-dotted line), perpendicular to the 
tangent at angle $\alpha$ (solid line) that intersects a curvi-linear 
feature (thick solid curve) at position $(x_0, y_0)$. The angle $\gamma$
indicates the half angular range of the curved guiding segment
(thick solid line).}
\end{figure}

\underbar{(6) Local curvature radius:} Next we determine the second-order 
term of a polynomial that follows the loop to be traced. We define a 
directional angle $\beta_0$ that is in perpendicular direction to the loop 
and defines the direction where all centers of  possible curvature radii 
are located (that intersect the structure at location $(x_0,y_0)$), see
geometric definition of the angles $\alpha$ and $\beta$ in Fig.~7,
\begin{equation}
	\beta_0 = \alpha_0  + {\pi \over 2} \ .
\end{equation}
The location $(x_c,y_c)$ of the curvature center with a minimum curvature 
radius $r_{min}$ is then found at,
\begin{equation}
	x_c = x_0 + r_{min} \cos{ \beta_0 } \ ,
\end{equation}
\begin{equation}
	y_c = y_c + r_{min} \sin{ \beta_0 } \ .
\end{equation}
The loci of all curvature centers $(x_m, y_m)$ of a set of curvature radii 
$r_m=r_{min}/[-1+2 m/(n_r-1)]$ (Eq.~6) is then found at,
\begin{equation}
	x_m = x_0 + (x_c - x_0) {r_m \over r_{min}} \ ,
\end{equation}
\begin{equation}
	y_m = y_0 + (y_c - y_0) {r_m \over r_{min}} \ .
\end{equation}
Since we want to follow a loop along a curved segment for every possible 
curvature radius $r_m$, we determine the coordinates for each segment 
point $s_k$. It is useful to define the angle $\beta_m$ of the line that 
connects a curvature center $(x_m, y_m)$ with a curved segment point $s_k$, 
\begin{equation}
	\beta_m = \beta_0 + \sigma_{dir} \left({ s_k \over r_m} \right) \ ,
\end{equation}
where $\sigma_{dir}=\pm 1$ has two opposite signs, depending on the 
forward or backward tracing of a loop.  The x,y-coordinates $(x_{km}, y_{km})$ 
of the loop segment $s_k$ is then,
\begin{equation}
	x_{km} =  x_m - r_m \cos{( \beta_m )} \ ,
\end{equation}
\begin{equation}
	y_{km} = y_m - r_m \sin{( \beta_m )} \ .
\end{equation}
In order to determine the curvature radius $r_m$ that fits the local loop 
segment best, we search for the segment with the maximum flux along the 
curve with radius $r_m$,
\begin{equation}
	z_{max} = max_m[ {1 \over n_r} \sum_{k=0}^{n_s-1} 
	z_{i,j}'(x_{km}, y_{km}) ] \ .
\end{equation}
Since we know now the optimum curvature radius $r_m$, we can trace the 
loop incrementally by a step $\Delta s$ and extrapolate the position 
$(x_{k+1}, y_{k+1})$ 
and angle $\alpha_{k+1}$ by
\begin{equation}
	\alpha_{k+1}=\alpha_k + \sigma_{dir} {\Delta s \over r_m} \ ,
\end{equation}
\begin{equation}
	\alpha_{mid} = {(\alpha_k + \alpha_{k+1}) \over 2} \ ,
\end{equation}
\begin{equation}
	x_{k+1} = x_k + \Delta s \cos{\left[ \alpha_{mid} + (\pi/2) 
	(1+\sigma_{dir}) \right]} \ ,
\end{equation}
\begin{equation}
	y_{k+1} = y_k + \Delta s \sin {\left[ \alpha_{mid} + (\pi/2)
	(1+\sigma_{dir}) \right]} \ .
\end{equation}

\begin{figure}
\centerline{\includegraphics[width=1.0\textwidth]{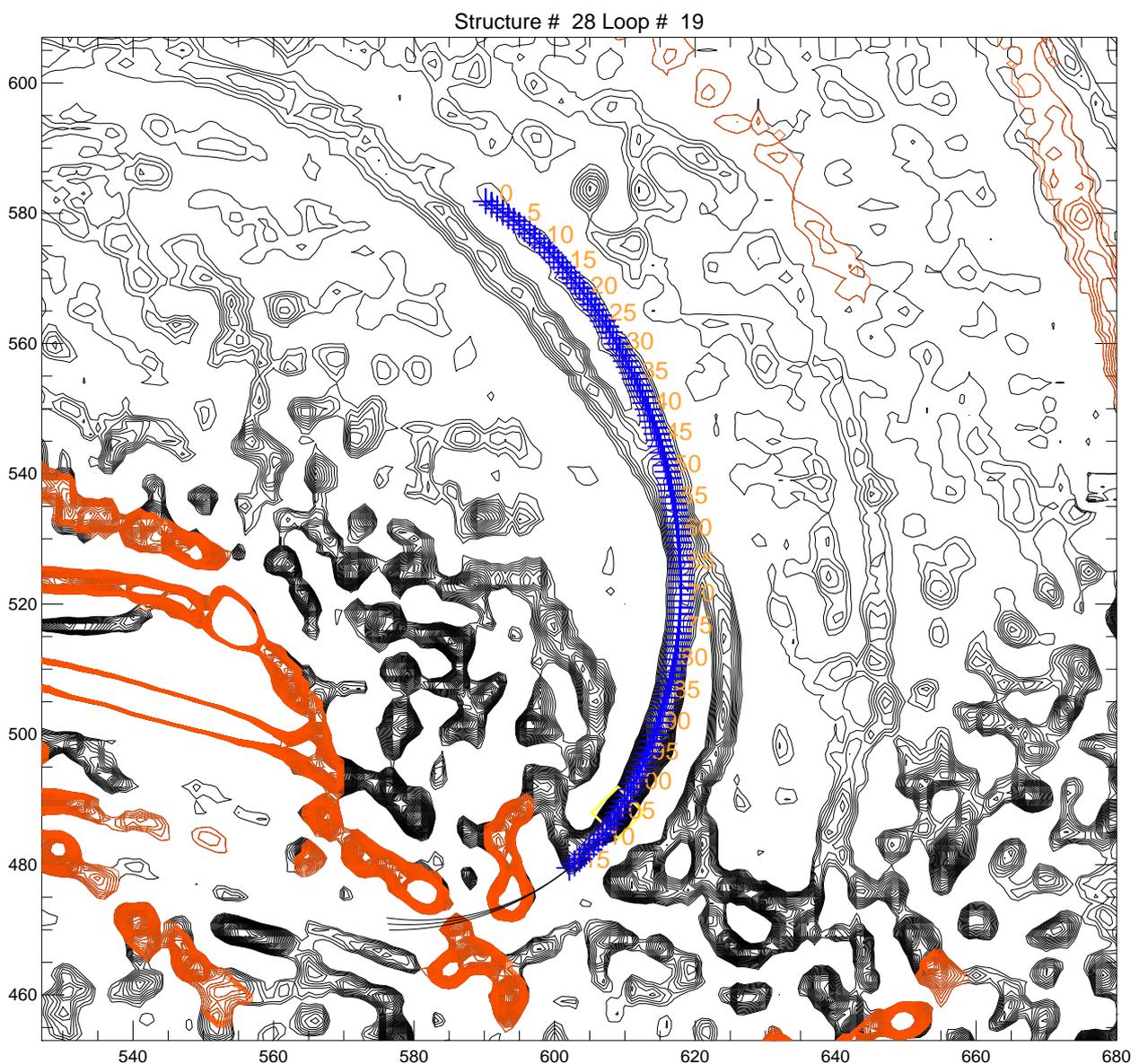}}
\caption{Example of loop tracing in the pixel area [525:680, 453:607]
of the image shown in Fig.~2. Loop \#19 is traced (blue crosses) over
a length of 115 pixels (orange numbers), crossing another structure at
a small angle. The curves at position 115 indicate the three curved
segments that have been used in the tracing of the last loop point.
The black contours indicate the bandpass-fitlered difference image,
and the red contours indicate the previously traced and erased
structures in the residual difference image.}
\end{figure}

\underbar{(7) Bidirectional tracing:} The step (6) describes the 
extrapolation or tracing of the loop segment from position $s_k=(x_k, y_k)$ 
to $s_{k+1}(x_{k+1},y_{k+1})$ by an incremental length step $\Delta s$. 
This step is repeated, starting from some arbitrary starting point $s_0$ 
inside the segment until the first endpoint $s_{n_{s1}}$, where the traced 
curvi-linear structure seems to end. The first endpoint generally is 
demarcated at a location where the bandpass-filtered image has zero or 
negative values, so one could just detect the first image pixel along 
the guiding segment $s_k$ that is non-positive. However, in order to allow for 
some minor gaps in noisy structures, it turned out to be more reliable 
to define a stop criterion when a least a few pixels have a nonpositive 
value, say $n_{gap}=3$ pixels. An example of a traced loop is shown
in Fig.~8, which illustrates that a loop can reliably be traced even
in the presence of secondary loops that intersect at small angles.

After completing the first half segment ($\sigma_{dir}=+1$), we repeat 
the same procedure from the midpoint $(x_0,y_0)$ in opposite direction
($\sigma_{dir}=-1$), 
until we stop at the second endpoint at $s_{n_{s2}}$. We combine than 
the two segments $[s_0, s_{n1}]$ and $[s_0,s_{n2}]$ by reversing one 
segment in order to maintain the same direction, and concatenating 
the two equal-directed segments into a single loop structure with 
indices $[s_0,...,s_{n_s}]$, where the length is $s_n=(s_{n_1}+s_{n2})$. 

\begin{figure}
\centerline{\includegraphics[width=1.0\textwidth]{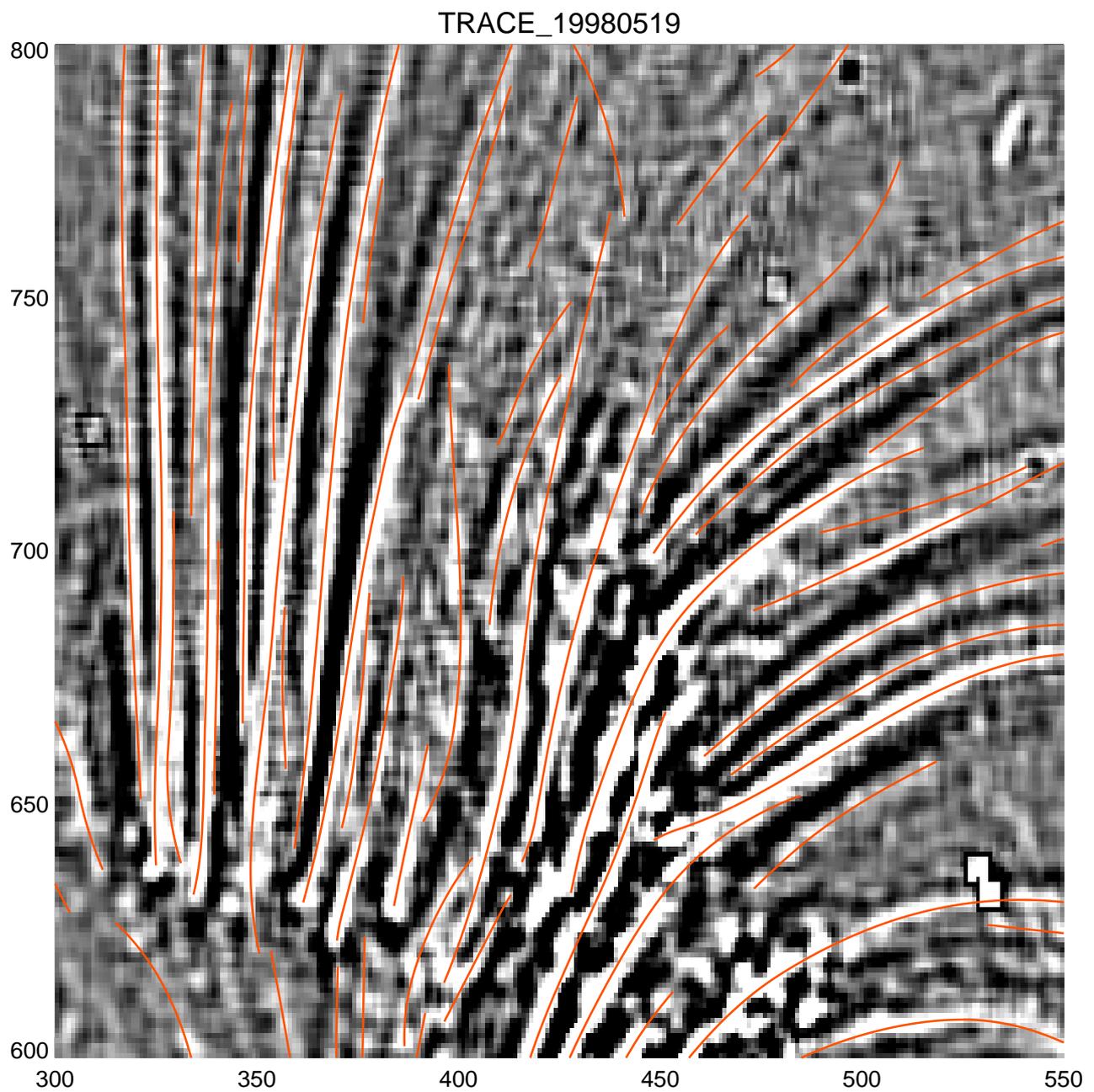}}
\caption{Example of loop tracings in area [300:550, 600:800] of the full 
image shown in Fig.~2. The greyscale indicates the bandpass-filtered
image $(n_{sm1}=5, n_{sm2}=7)$, and the loop tracings are shown with red
curves.}
\end{figure}

\underbar{(8) Loop Iteration:} The steps (4) to (7) describe the tracing 
of a single loop, specified by the $n_L$ coordinates 
$(x_i,y_i)=[x(s_k),y(s_k)], k=0,...,n_L-1$. 
In order to proceed to the next loop we want first to erase 
the area of this traced loop in the residual image, so that no previously 
traced loop segment is traced a second time in a consecutive loop. 
The loop is erased in the residual image simply by setting those 
image pixels to zero that have a distance of $d \le n_w =(n_{sm2}/2-1)$ 
from the 
loop coordinates, i.e., $z^{filter}[x_i\pm n_w, y_j \pm n_w]=0$. 
An example of an image area with dense coverage of loops is shown
in Fig.~9, where some loops are traced at flux levels close to the 
noise threshold. 

\underbar{(9) Loop Parameters:} After we described analytically the 
numerical code, we summarize the free parameters and the dependent 
parameters (that do not require any selection by the user of the code). 
The code has the following free or input parameters: the lowpass 
filter boxcar $n_{sm1}$, the minimum curvature radius $r_{min}$, 
the base level factor $q_{med}$, in units of image median brightness 
values, and the corner coordinates of the noise area $[i_{n1}:i_{n2}, 
j_{n1}:j_{n2}]$. All other parameters 
are dependent, or a fixed constant, such as the tracing step $\Delta s=1$, 
the highpass filter boxcar $n_{sm2}=n_{sm1}+2$, 
the half width of the erased loop cross-sections $n_w=n_{nsm2}/2-1$, 
and the loop termination gap $n_{gap}=3$.

\acknowledgements{Acknowledgements}
We thank Ilya Grigoriev and Anna Akhmanova for providing the images
of labelled microtubules. We acknowledge also helpful discussions with
Karel Schrijver, Mark DeRosa, Allen Gary, Jake Lee, Bernd Inhester,
James McAteer, Peter Gallagher, Alex Young, Alex Engels, Patrick Shami,
Narges Fathalian, Fatemeh Amirkhanlou, Hossein Safari, and participants
of the {\sl 3rd Solar Image Processing Workshop} in Dublin, Ireland,
$6-8$ September 2006, the {\sl 4th Solar Image Processing Workshop}
in Baltimore, Maryland, $26-30$ October 2008, the {\sl 5th
Solar Imaging Processing Workshop} in Les Diablerets, Switzerland,
$13-16$ September 2010, and the {\sl 6th Solar Imaging Processing Workshop}
at Montana State University, Bozeman, $13-16$ August 2012.
Part of the work was supported by the NASA TRACE contract (NAS5-38099) 
and the NASA SDO/AIA contract NNG04EA00C.

\bibliographystyle{mdpi}
\makeatletter
\renewcommand\@biblabel[1]{#1. }
\makeatother


\end{document}